\documentclass{appolb}
\usepackage{graphicx}
\usepackage{ogonek}
\usepackage{amsmath}

\newcommand{\gras}[1]{\boldsymbol{#1}}

\begin{document}
\title{Mirror and triplet displacement energies\\within nuclear DFT: numerical stability
\thanks{Presented at the Zakopane Conference on Nuclear Physics “Extremes of the Nuclear
Landscape”, Zakopane, Poland, August 28 -- September 4, 2016}%
}

\author{P.~B\k{a}czyk$^1$, J.~Dobaczewski$^{1,2,3,4}$, M.~Konieczka$^1$, T.~Nakatsukasa$^{5,6}$, K. Sato$^7$, W.~Satu\l{}a$^{1,4}$
\address{$^1$Faculty of Physics, University of Warsaw, Pasteura 5, PL-02-093 Warsaw, Poland\\
$^2$Department of Physics, University of York, Heslington, York YO10 5DD, UK\\
$^3$Department of Physics, P.O. Box 35 (YFL), FI-40014  University of Jyv\"askyl\"a, Finland\\
$^4$Helsinki Institute of Physics, P.O. Box 64, FI-00014 University of Helsinki, Finland\\
$^5$ Center for Computational Sciences, University of Tsukuba, Tsukuba 305-8577, Japan\\
$^6$ RIKEN Nishina Center, Wako 351-0198, Japan\\
$^7$ Department of Physics, Osaka City University, Osaka 558-8585, Japan}
}

\maketitle
\begin{abstract}
Isospin-symmetry-violating class II and III contact terms are introduced into the Skyrme energy density functional to account for charge dependence
of the strong nuclear interaction. The two new coupling constants are adjusted to available experimental data on triplet and mirror
displacement energies, respectively.  We present preliminary results of the fit, focusing on its numerical stability with respect to
the basis size.
\end{abstract}

\section{Introduction}
An accurate description of atomic nucleus, a system of protons and neutrons interacting with electromagnetic and strong forces, is a difficult task.
It can be simplified considerably by introducing the concept of isospin symmetry  \cite{[Wig37]} that relies on charge independence, that is, on equality
of nucleon-nucleon (NN) forces $V_{pp}=V_{pn}=V_{nn}$ in the same space-spin channel.
The NN scattering experiments indicate, however, that the strong interaction depends slightly on a pair of nucleons involved
in the process \cite{[Mil90]}.  On a fundamental level, violation of the isospin symmetry is due to the mass splitting and different
charges of the $up$ and $down$ quarks and the difference in quark composition of proton and neutron.

In atomic nuclei, the main source of isospin-symmetry breaking (ISB) is the Coulomb force, which shifts binding energies of nuclei forming a multiplet of a given isospin $T$.
This property is used to construct various mass indicators which are sensitive to the ISB effects. The most common among such indicators are
the mirror (MDE) and triplet (TDE) displacement energies involving data on isospin doublets ($T=\frac12$) and triplets ($T=1$).
It turns out, however, that the Coulomb interaction alone is not sufficient to fully explain neither the MDEs nor the TDEs and the additional ISB
mechanism due to the strong nuclear force might be of importance in the understanding of the experimental data \cite{[Nol69],[Sat14s]}.

Contemporary \textit{ab initio} models are able to account for ISB effects in both NN scattering data and light nuclei \cite{[Wir95],[Mac01a],[Wir13s]}. However, they are still not suitable for describing heavier systems which is a domain of mean-field (MF) or density functional theory (DFT). These approaches are excellent tools to study bulk properties (masses, radii or quadrupole moments) in atomic nuclei  regardless of their mass and parity
of proton and neutron numbers, see~\cite{[Ben03]} and references therein. Among different variants of MF or DFT approaches the models based on the Skyrme interaction~\cite{[Sky59]} are the most efficient computationally and fairly well describe nuclear binding energies.
 However, the isospin invariant Skyrme energy density functionals (S-EDF) \cite{[Sat13],[She14]}, which are typically used in practical applications,
 systematically fail to reproduce the experimental data on MDEs and TDEs. In our recent work \cite{[Bac15]} we introduced two new ISB terms into
 the S-EDF. They read:
\begin{gather}
\hat{V}^{{II}}(i,j) = \frac12 t_0^{\rm{II}}\, \delta\left(\gras{r}_i - \gras{r}_j\right)
\left(1 - x_0^{\rm{II}}\,\hat P^\sigma_{ij}\right) \times \left[3\hat{\tau}_3(i)\hat{\tau}_3(j)-\hat{\vec{\tau}}(i)\circ\hat{\vec{\tau}}(j)\right],\\
\hat{V}^{\rm{III}}(i,j) = \frac12 t_0^{\rm{III}}\, \delta\left(\gras{r}_i - \gras{r}_j\right)\left(1 - x_0^{\rm{III}}\,\hat P^\sigma_{ij}\right) \times \left[\hat{\tau}_3(i)+\hat{\tau}_3(j)\right].
\end{gather}
 The first calculations performed with these modifications proved the ability of the extended model to correctly grasp the missing
 ISB effects in both the MDEs and TDEs, however, a systematic fitting of the new coupling constants $t_0^\mathrm{II}$ and $t_0^\mathrm{III}$ is necessary ($x_0^\mathrm{II}$ and $x_0^\mathrm{III}$ turned out to be redundant).

In this paper we present a discussion on the numerical stability of MDEs and TDEs with respect to the choice of the basis size. The
study allows us to estimate a theoretical uncertainty associated with a given basis cut-off.  The discussion is followed by a presentation
of  preliminary results of the fitting procedure performed with a certain basis cut-off.

\section{Numerical stability}\label{sec:stab}

Nuclear calculations often depend on a choice of the basis size. Their reliability and, in particular, predictive power requires an estimate of theoretical uncertainties related to the basis-size. The HFODD code \cite{[Sch12a]} used  in this work solves the HF equation in the Cartesian harmonic oscillator (HO)  basis. Its size can be controlled  by providing a number of spherical HO
shells $N$. In practical applications, the choice of $N$  is always a matter of  trade-off between computation time and
expected precision of the calculations.

\begin{figure}[htb]
\centerline{%
\includegraphics[scale=0.83]{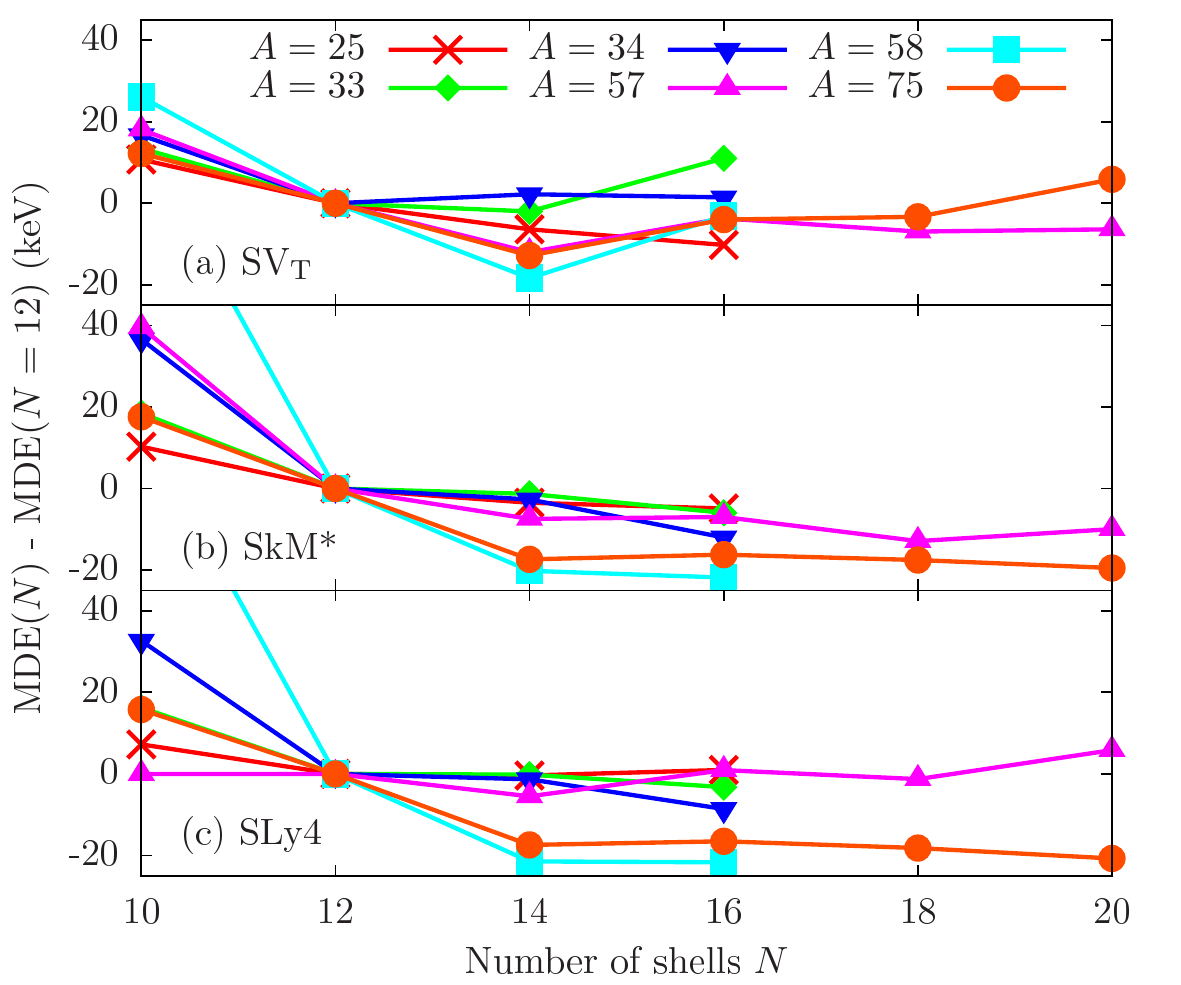}}
\caption{(Color online) Values of MDEs in function of the number of spherical HO shells used in the basis, plotted with respect to that
obtained for $N=12$. Results for multiplets with different values of $A$ are labelled with different symbols, as shown in the legend.
Panels (a), (b), and (c) show values obtained using SV$_\mathrm{T}$, SkM*, and SLy4 S-EDFs, respectively.}
\label{fig:MDE}
\end{figure}

\begin{figure}[htb]
\centerline{%
\includegraphics[scale=0.83]{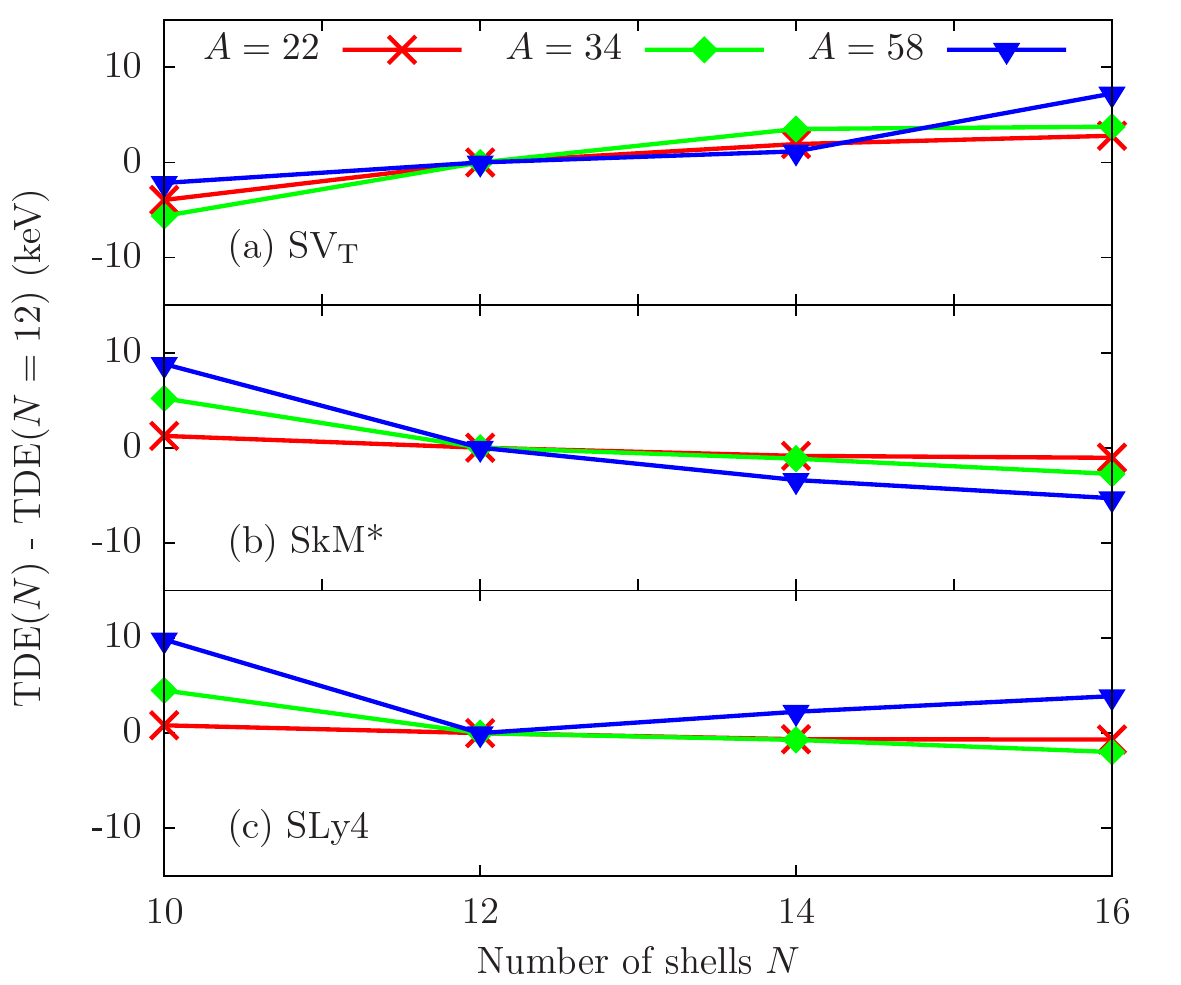}}
\caption{(Color online) Same as in Fig.~\protect\ref{fig:MDE}, but for the TDEs.}
\label{fig:TDE}
\end{figure}

To evaluate theoretical uncertainty of the calculated MDEs and TDEs due to  the basis size we have performed  test calculations for $T=\frac12$ doublets with $A=25$, 33, 57, and 75 and for $T=1$ triplets with $A=22$, 34, and 58. Calculations have been performed using the spherical HO bases consisting of $N=10$, 12, 14, and 16 shells. For the heaviest doublets with $A=57$ and 75, we have extended the test by including the
bases consisting of $N=18$ and 20 shells. In each case we have computed the MDEs and TDEs using three different S-EDF
parametrizations, including the density-independent SV$_{\rm T}$ S-EDF of Refs.~\cite{[Bei75],[Sat11]} and two popular density-dependent S-EDFs
SkM$^*$ \cite{[Bar82]} and SLy4~\cite{[Cha98]}. Additionally, it has been checked that the new ISB terms affect rather weakly the stability of the MDEs and the TDEs and, in effect, the calculations were performed with $t_0^\mathrm{II}$ and $t_0^\mathrm{III}$ as in Table~\ref{tab:t-param}. Results are collected in Figs.~\ref{fig:MDE} and \ref{fig:TDE}.

The basis-size-dependence tests suggest that the optimal strategy regarding both efficiency and accuracy of the calculations
is to compute  light  ($10 \leq A \leq 30$),  medium-mass  nuclei ($31 \leq A \leq 56$),
and heavy  ($A\geq 57$) nuclei using bases consisting of $N=10$, 12, and 14 spherical HO shells, respectively.
Both for the MDEs and TDEs, this strategy would result in the basis-size related uncertainty not exceeding $\Delta_{{\rm basis}}\approx \pm$15 keV.
As we show in Sect.~\ref{sec:Fit}, the basis-size uncertainty is relatively small as compared to the uncertainty resulting from the
fitting procedure, which may justify using even smaller bases.

\section{Results of fitting}\label{sec:Fit}

As stated in Sec.~\ref{sec:stab}, the optimum choice of the number of spherical HO shells implies dividing the mass region of interest into
the three subsets.
In the present preliminary study, we use $N$=10 shells for all nuclei having masses $10 \leq A \leq 56$ and 58,
and we use $N=14$ shells only for the heaviest systems. This choice almost doubles the basis-size related uncertainty of the calculated MDEs
to $\Delta_{{\rm basis}}\approx\pm$30\,keV without much affecting the uncertainty of the calculated TDEs. A smaller computational time
allows us to explore the richness of experimental data available for ISB effects in the $10 \leq A \leq 58$ nuclei. In future, it may
allow us to perform calculations of mirror and triplet energy differences in rotational bands or
the ISB effects in electromagnetic and  $\beta$ decays using the recently developed no-core configuration-interaction (NCCI) formalism
\cite{[Sat16]}, which involves CPU time demanding isospin and angular momentum projections and configuration mixing.

With the less strict choice of the basis size, the two new coupling
constants (see Ref.~\cite{[Bac15]}) were adjusted to all available
data on MDEs ($10 \leq A \leq 75$) and TDEs ($10 \leq A \leq 58$). The fitting procedure
has been realized independently for $t_0^{\mathrm{II}}$ and
$t_0^{\mathrm{III}}$ parameters using linear regression method and
following the guidelines from Ref.~\cite{[Dob14]} and will be
described in details in our forthcoming publication. The fitting
results as well as the standard deviations from the experimental data
points are presented in Table~\ref{tab:t-param}.

\begin{table}[h!]
\centering
\caption{Values of the $t_0^{\mathrm{II}}$ and $t_0^{\mathrm{III}}$ parameters with uncertainty
and standard deviations for MDE and TDE, $\sigma_{\rm fit}$, resulting from the fit to all available
data on isospin doublets and triplets.
The calculations has been done for three different S-EDFs considered in this work.}
\label{tab:t-param}
\medskip
\begin{tabular}{|c|c|c|c|}\hline
Interaction & SV & SkM* & SLy4\\\hline
$t_0^{\mathrm{II}}$ (MeV fm$^3$)& $17\pm5$ & $24 \pm 8$ & $22 \pm 7$\\
$\sigma_{\rm fit}$ (keV) & 100 & 110 & 100\\ \hline
$t_0^{\mathrm{III}}$ (MeV fm$^3$)& $-7.3\pm1.9$ & $-5.5 \pm 1.3$ & $-5.5 \pm 1.1$\\
$\sigma_{\rm fit}$ (keV) & 190 & 150 & 120 \\ \hline
\end{tabular}
\end{table}

As it turns out, the uncertainty resulting from the choice of the basis cut-off, $ \Delta_{{\rm basis}}$, is small with respect to the values
of standard deviations,  $ \sigma_{{\rm fit}}$,  given in Table~\ref{tab:t-param}. Hence, in the total uncertainty,
$\Delta_{\rm T} = \sqrt{\sigma_{{\rm fit}}^2 + \Delta_{{\rm basis}}^2}$, the basis-size related uncertainty constitutes only a small correction.

\section{Summary}

In the paper we performed preliminary calculations of MDEs and TDEs
paying special attention to the numerical stability of the results
with respect to the basis size. It turned out that the optimum choice of
the number of spherical HO shells, $N$, which defines the basis size,
is $N=10$ for light nuclei ($10 \leq A \leq 30$), $N=12$ for medium-mass
nuclei ($31 \leq A \leq 56$) and $N=14$ for heavy nuclei ($A\geq 57$). It is
shown, that the corresponding uncertainty, $\Delta_{\rm basis}
= \pm 15$\.keV is small as compared to the uncertainty coming from
the fitting procedure. This allows us to compute light and
medium-mass nuclei using the basis consisting only of $N=10$ HO shells
without loosing much of the precision. Smaller basis, in turn,  will be
beneficial regarding efficiency of the planned NCCI calculations.
Finally, the two new coupling constants were adjusted to all available
experimental data. The details of the fitting procedure will be
presented in the forthcoming publications.

This work has been supported by the Polish National Science Centre
under the contract 2015/17/N/ST2/04025 and by the
Academy of Finland and University of Jyv\"askyl\"a within the FIDIPRO
program. Advanced computational
resources were provided by the \'{S}wierk Computing Centre (CIS) at
the National Centre for Nuclear Research (NCBJ) and CSC-IT Centre for Science Ltd, Finland.


\begin{thebibliography}{99}

\bibitem{[Wig37]}{E. Wigner, \textit{Phys. Rev.} {\bf 51}, 106 (1937)}.

\bibitem{[Mil90]}{G.A. Miller, B.M.K. Nefkens, and I. \v{S}laus, \textit{Phys. Rep.} {\bf 194}, 1 (1990)}.

\bibitem{[Nol69]}{J.A. Nolen and J.P. Schiffer, \textit{Ann. Rev. Nuc. Sci.} {\bf 19}, 471 (1969)}.

\bibitem{[Sat14s]}{W. Satu\l{}a, J. Dobaczewski, M. Konieczka, and W. Nazarewicz, \textit{Acta Phys. Polon. B} {\bf 45}, 167 (2014).}

\bibitem{[Wir95]}{R.B. Wiringa, V. G. J. Stoks, and R. Schiavilla, \textit{Phys. Rev. C} {\bf 51}, 38 (1995)}.

\bibitem{[Mac01a]}{R. Machleidt, \textit{Phys. Rev. C} {\bf 63}, 024001 (2001)}.

\bibitem{[Wir13s]}{R.B. Wiringa, S. Pastore, S.C. Pieper, and G.A. Miller, \textit{Phys. Rev. C} {\bf 88}, 044333 (2013).}

\bibitem{[Sky59]}{T.H.R. Skyrme, \textit{Nucl. Phys.} {\bf 9}, 615 (1959)}.

\bibitem{[Ben03]}{M. Bender, P.-H. Heenen, and P.-G. Reinhard, \textit{Rev. Mod. Phys.} {\bf 75}, 121 (2003)}.

\bibitem{[Sat13]}{K. Sato, J. Dobaczewski, T. Nakatsukasa, and W. Satu\l{}a, \textit{Phys. Rev. C} {\bf 88}, 061301(R) (2013)}.

\bibitem{[She14]}{J.A. Sheikh, {\it et al.}, \textit{Phys. Rev. C} {\bf 89}, 054317 (2014)}.

\bibitem{[Bac15]} {P. B\k{a}czyk, J. Dobaczewski, M. Konieczka, and W. Satu{\l}a, \textit{Acta Phys. Pol. B Proc. Suppl.} {\bf 8}, 539 (2015)}.

\bibitem{[Sch12a]}{N. Schunck {\it et al.}, \textit{Comput. Phys. Commun.} {\bf 183}, 166 (2012)}.

\bibitem{[Bei75]}{M. Beiner, H. Flocard, N. Van Giai, and P. Quentin, \textit{Nucl. Phys. A} {\bf 238}, 29 (1975)}.

\bibitem{[Sat11]}{W. Satu{\l}a, J. Dobaczewski, W. Nazarewicz, and M. Rafalski, \textit{Phys. Rev. Lett.} {\bf 106}, 132502 (2011)}.

\bibitem{[Bar82]}{J. Bartel, {\it et al.}, \textit{Nucl. Phys. A} {\bf 386}, 79 (1982)}.

\bibitem{[Cha98]}{E. Chabanat, {\it et al.}, \textit{Nucl. Phys. A} {\bf 635}, 231 (1998)}.

\bibitem{[Sat16]}{W. Satu{\l}a, P. B\k{a}czyk, J. Dobaczewski, and M. Konieczka, \textit{Phys. Rev. C} {\bf 94}, 024306 (2016)}.

\bibitem{[Dob14]}{J. Dobaczewski, W. Nazarewicz, and P.G. Reinhard, \textit{J. Phys. G: Nucl. and Part. Phys.} {\bf 41}, 074001 (2014)}.

%
%
%
%
%
%
%
%
%
%
%
%
%
%
%
%
%
%

\end{thebibliography}
\end{document}